\documentclass[11pt,twoside]{article}

\usepackage{asp2004}

\usepackage{epsf}

\usepackage{psfig}

\usepackage{lscape}
\usepackage{graphicx}

\begin{document}

\title{Investigation of the magnetic field characteristics of Herbig Ae/Be stars: Discovery of the pre-main sequence progenitors of the magnetic Ap/Bp stars}   

\author{G.A. Wade$^1$, D. Drouin$^1$, S. Bagnulo$^2$, J.D. Landstreet$^3$, E. Mason$^2$, J. Silvester$^{1,4}$, E. Alecian$^5$, T. B\"ohm$^6$, J.-C. Bouret$^7$, C. Catala$^5$, J.-F. Donati$^6$, C. Folsom$^{1,4}$, K. Bale$^1$}   

\affil{$^1$Department of Physics, Royal Military College of Canada, Kingston, Canada}    
\affil{
   $^2$European Southern Observatory, Santiago, Chile}
\affil{
   $^3$Department of Physics \& Astronomy, University of Western Ontario, London, Canada}
\affil{
   $^4$Department of Physics, Queen's University, Kingston, Canada}
\affil{
   $^5$Obs. de Paris LESIA, Meudon, France}
\affil{
   $^6$Obs. Midi-Pyr\'en\'ees, Toulouse, France}
\affil{
   $^7$Laboratoire d'Astrophysique de Marseille, Marseille, France}

\begin{abstract}
We are investigating the magnetic characteristics of pre-main sequence Herbig Ae/Be stars, with the aim of (1) understanding the origin and evolution of magnetism in intermediate-mass stars, and (2) exploring the influence of magnetic fields on accretion, rotation and mass-loss at the early stages of evolution of A, B and O stars. We have begun by conducting 2 large surveys of Herbig Ae/Be stars, searching for direct evidence of photospheric magnetic fields via the longitudinal Zeeman effect. From observations obtained using FORS1 at the ESO-VLT and ESPaDOnS at the Canada-France-Hawaii Telescope, we report the confirmed detection of magnetic fields in 4 pre-main sequence A- and B-type stars, and the apparent (but as yet unconfirmed) detection of fields in 2 other such stars. We do not confirm the detection of magnetic fields in several stars reported by other authors to be magnetic: HD 139614, HD 144432 or HD 31649. One of the most evolved stars in the detected sample, HD~72106A, shows clear evidence of strong photospheric chemical peculiarity, whereas many of the other (less evolved) stars do not. The magnetic fields that we detect appear to have surface intensities of order 1 kG, seem to be structured on global scales, and appear in about 10\% of the stars studied. Based on these properties, these magnetic stars appear to be pre-main sequence progenitors of the magnetic Ap/Bp stars. 

\end{abstract}









\section{Introduction}

Ap/Bp stars represent about 5\% of all intermediate-mass main sequence stars and are characterized by strong, globally-ordered surface magnetic fields. The fundamental origin of these magnetic fields remains fundamentally a mystery. Because A and B type stars lack an outer convective envelope and exhibit none of the characteristics we have come to associate with dynamo-generated fields, it is generally believed that the magnetic fields of Ap stars are remnants, or fossils, of the interstellar magnetic field swept up during star formation. 


A natural consequence of this ``fossil field'' hypothesis is the existence of magnetic field in some pre-main sequence (PMS) stars of intermediate-mass - stars which would become Ap/Bp stars when they reach the main sequence. Furthermore, we expect that the fields of those PMS stars will exhibit similar incidence, structure and intensity to those we observe in the Ap stars. The goals of our surveys were to search for such fields in the Herbig Ae/Be (HAeBe) stars, which are PMS stars of intermediate mass, of spectral types A and B, and which are characterised spectroscopically by the presence of strong, and often variable emission lines.

\section{Experiment}

More than 50 HAeBe stars were surveyed for circular polarization due to the longitudinal Zeeman effect in their absorption lines. This task was performed using the FORS1 low-resolution ($R\sim 1950$) spectropolarimeter at the ESO Very Large Telescope (VLT) and the ESPaDOnS high-resolution ($R\sim 65000$) spectropolarimeter at the Canada-France-Hawaii Telescope (CFHT). The data obtained at the VLT were analyzed using regression methods as described by Bagnulo et al. (2002). On the other hand, the data gathered at the CFHT were analyzed using the Least-Squares Deconvolution multi-line analysis procedure (Donati et al. 1997).

The Stokes $V$ signatures and longitudinal magnetic field diagnosed from the low-resolution polarised spectra obtained with FORS1 are virtually insensitive to the projected rotational velocity of the star. This has the advantage of providing a $v\sin i$-unbiased assessment of the magnetic characteristics of HAeBe stars, and allowing us to obtain relatively uniform longitudinal field error bars (typically about 50~G). The Stokes $V$ signatures and longitudinal fields diagnosed from the high-resolution ESPaDOnS spectra are highly sensitive to $v\sin i$, but the high-resolution spectra provide the ability to resolve often complex line profiles into their individual absorption and emission components, sometimes allowing discovery of polarisation which was undetected in the low-resolution spectra. Together, the low- and high-resolution strategies are mutually complementary.

The positions of the selected stars on the HR diagram (derived primarily using temperatures and luminosities available in the literature, along with the theoretical PMS evolutionary tracks of Palla \& Stahler 1993) indicate that most of these stars have masses between 2 and $3~M_\odot$. Using this property of the sample, combined with the incidence statistics of Ap stars (Wolff 1968, Johnson 2004), we estimate that about 10\% of the observed HAeBe stars will eventually become Ap stars and therefore should host magnetic fields according to the fossil hypothesis. From our observations and analysis we have confidently detected and confirmed magnetic fields in 4 of the HAeBe stars observed (HD 72106, V380 Ori, HD 190073 and HD 200775). We have apparently detected fields in 2 additional HAeBe stars (HD 101412 and BF Ori), although these detections have not yet been confirmed. Considering HD~104237 (a HAeBe star in which Donati et al. (1997) detected a field), there are now 5 confirmed magnetic HAeBe stars. A preliminary summary of these results was reported by Wade et al. (2005). The general characteristics of these stars are summarised in Table 1.

\begin{table}
\begin{center}
\begin{tabular}{lcccc}
\hline\hline
\noalign{\smallskip}
Object &  $T_{\rm eff}$ (K) & Mass ($M_\odot$) & Age (Myr) & $B_{\rm surf}$ (G)\\
\noalign{\smallskip}
\hline
\noalign{\smallskip}
\multicolumn{5}{c}{Confirmed}\\
\noalign{\smallskip}
\hline
\noalign{\smallskip}
HD 104237& 8500  & 2.3 & 2  & $>150$\\
HD 72106 & 11000 & 2.4 & 10 & $\sim 800$\\
V380 Ori & 10700 & 2.8 & 1 & $>1170$\\
HD 200775& 18000 & 10? & $<1$ & $>1800$\\
HD 190073&  9250 & 2.9 & $<1$  & $>120$\\
\noalign{\smallskip}
\hline
\noalign{\smallskip}
\multicolumn{5}{c}{Suspected}\\
\noalign{\smallskip}
\hline
\noalign{\smallskip}
HD 101412& 9500  & 2.6 & 2 & $>1065$\\
BF Ori   & 8750? & ?   & ? & $>370$\\
\noalign{\smallskip}\hline\hline
\end{tabular}
\caption{Basic properties of known and suspected magnetic HAeBe stars. Surface field ($B_{\rm surf}$) lower limits are estimated from small numbers of longitudinal field measurements, assuming a predominantly dipolar field. $B_{\rm surf}$ for HD 72106 is estimated from an approximate fit to the LSD Stokes $V$ profiles, whereas $B_{\rm surf}$ for HD 200775 is inferred from the phased longitudinal field variation.
}
\end{center}
\end{table}

Here we briefly discuss the properties of various sample stars (those both detected and undetected), the characteristics of the detected fields, and the observed field incidence.

\subsection{HD 72106}

HD 72106 is a visual double in which the magnetic field was detected with ESPaDOnS. The two components of this system are both A-type stars ($T_{\rm pri}=11000$~K and $T_{\rm sec}=8500$~K; Drouin et al., in preparation). They were observed together by Hipparcos, giving a consistent distance of 288 pc. They show no detectable relative proper motion, despite large absolute proper motion. Our data show no significant difference in radial velocity, and it appears that the two components are physically associated.

Magnetic field is detected in the primary component with ESPaDOnS. No field is detected in the secondary component, either with ESPaDOnS or with FORS1. The primary does not appear to show HAeBe characteristics, whereas the secondary is classified by Vieira et al. (2003) as an ``evolved HAeBe star'' and shows clear H$\alpha$ and He~{\sc i}~$\lambda 5876$ emission. The high-resolution ESPaDOnS spectra of the combined system and of the individual components show line profile variability, and indicates that the magnetic primary exhibits chemical peculiarity similar to that of Ap/Bp stars. Example Least-Squares Deconvolved Stokes $I$ and $V$ profiles of HD 72106 are shown in Fig. 1.

Five ESPaDOnS observations of the magnetic primary obtained so far show a changing surface field topology, indicating a rotational period of about 2 days - a value consistent with the observed $v\sin i$ and inferred radius. The morphology and modulation of the Stokes $V$ profiles indicate that the surface magnetic field is primarily dipolar, with an inferred polar strength of approximately 1.6~kG.

\begin{figure}
\centering
\includegraphics[width=6.0cm]{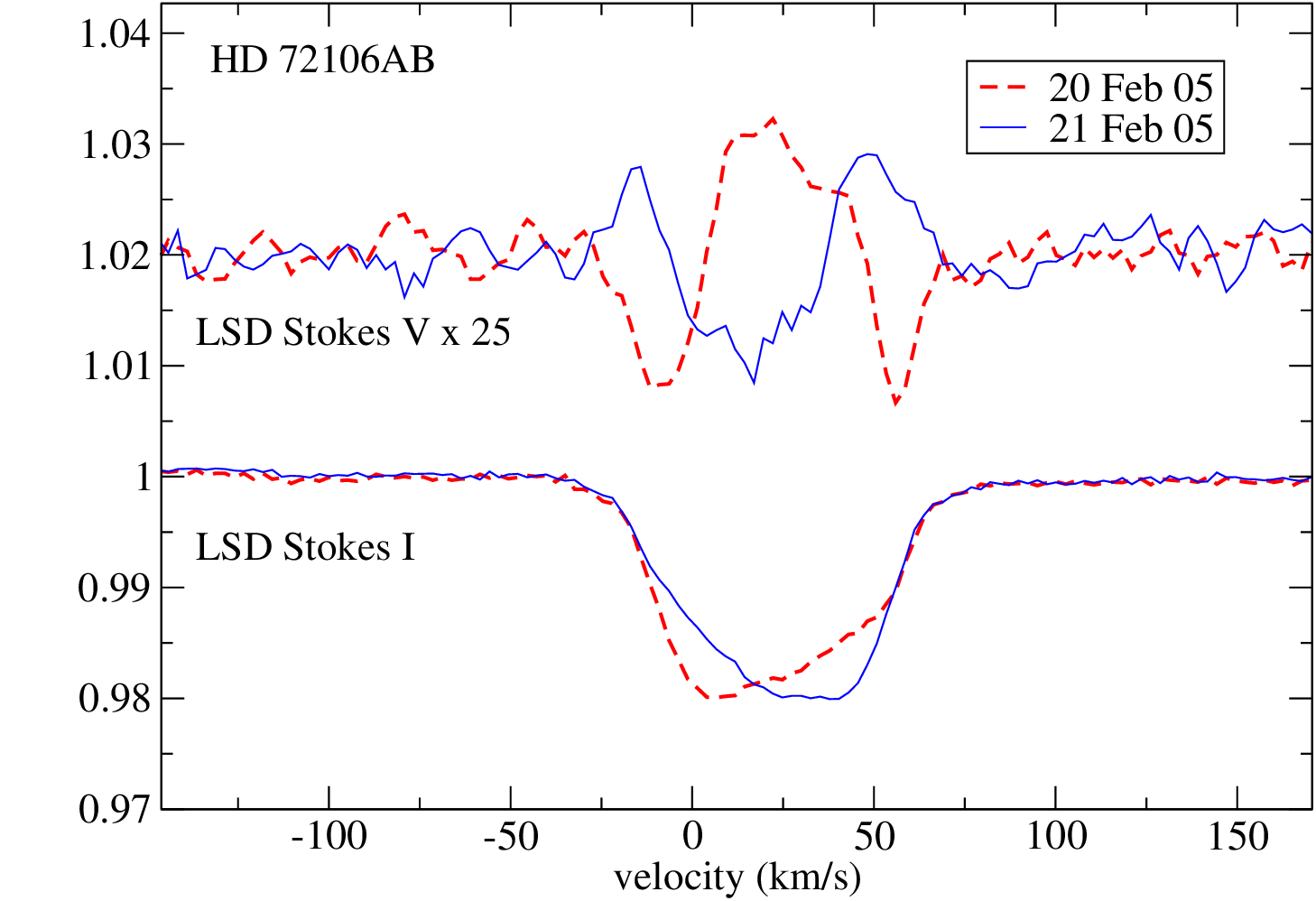}\hspace{0.5cm}\includegraphics[width=6.0cm]{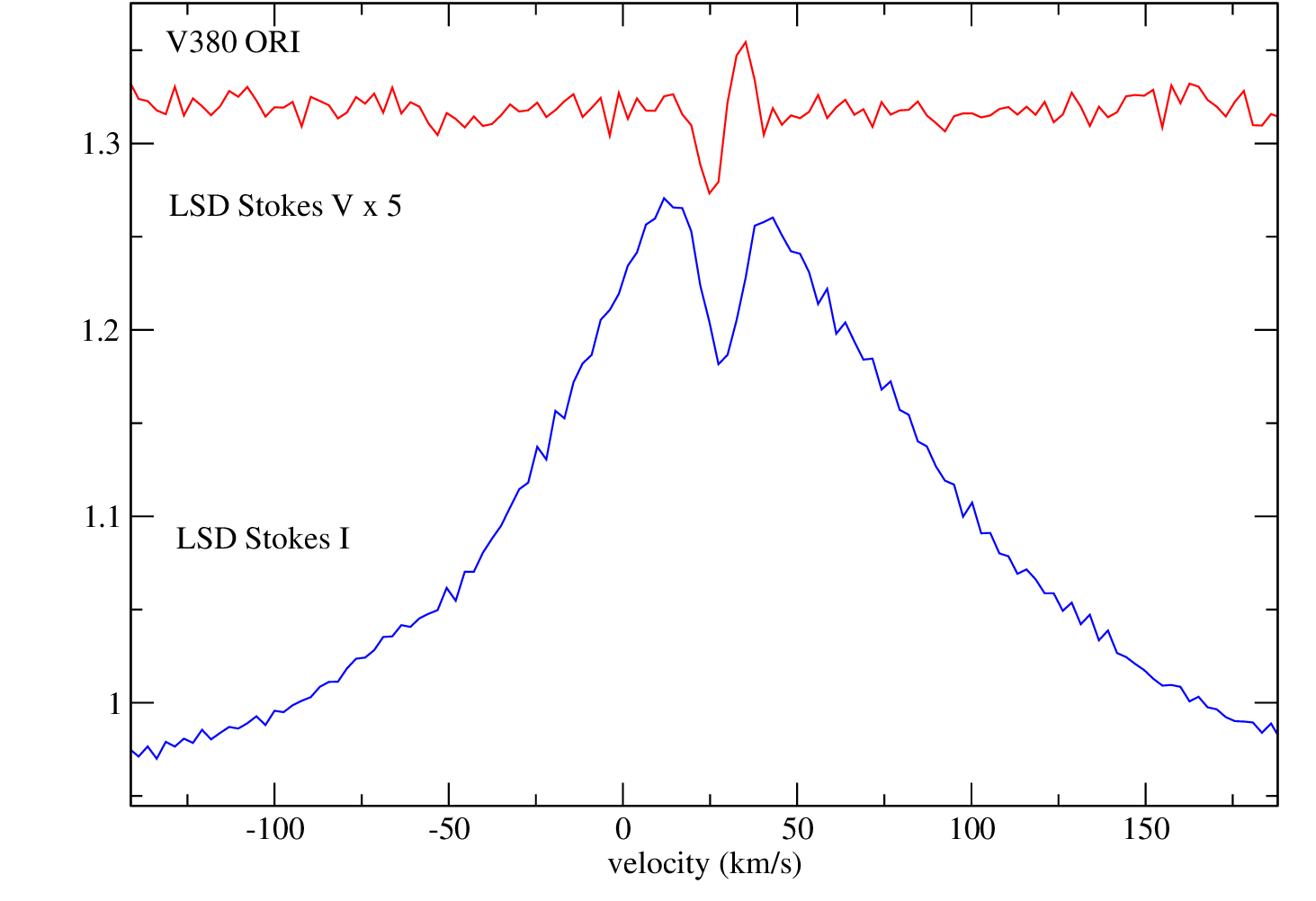}
\caption{Stokes $I$ and $V$ LSD profiles of HD 72106 (left) and V380 Ori (right). Note the clear Zeeman signatures associated with the absorption lines. Adapted from Wade et al. (2005).}
\end{figure}

\subsection{V380~Ori}

V380~Ori is a young B9 HAeBe star exhibiting numerous emission lines in its optical spectrum, associated with essentially all metallic, H and He transitions. Many lines have a weak, sharp central absorption component which appears to be at least partly photospheric in nature - it is in these absorption components that Stokes $V$ signatures are detected in the ESPaDOnS spectra. Rossi et al. (1999) have examined the photometric and spectroscopic properties of this star, concluding that it hosts a Keplerian disc and a cool wind. Leinert et al. (1997) discovered that V380 Ori is an interferometric binary, with a companion that is 3-4 times fainter in K band, has $v\sin i\sim 30$~km/s and which is sufficiently cool to display the Li~{\sc i}~$\lambda$6708 line in its spectrum (Corporon \& Lagrange 1999).

Four ESPaDOnS observations of V380 obtained so far indicate an approximately constant surface magnetic field topology, indicating that V380 Ori probably has a very long rotational period, is viewed nearly pole-on, or has magnetic and rotational axes that are nearly aligned. The inferred longitudinal field strength (nearly 0.5 kG) implies that the surface field is organised on large scales, with an intensity of about 1.5~kG if it is primarily dipolar. Inference of the absolute magnetic field intensity is particularly challenging due to the strong emission and intense monthly (and even nightly!) variability of this star. LSD Stokes $I$ and $V$ profiles are illustrated in Fig. 1. 


\subsection{HD~200775}

HD 200775 is a double-lined spectroscopic binary, with a B2 HAeBe primary and a later B or A-type secondary. Pogodin et al. (2004) determined preliminary orbital elements for the system, in particular the orbital period $P_{\rm orb}=1345$~d. Twelve ESPaDOnS spectra of this system have been obtained. The reported orbital period and the derived projected rotational velocities of the components (primary: 23~km/s; secondary: 65~km/s) indicate that HD 200775 is a detached, unsynchronised system.

According to Catala et al. (in preparation), magnetic field is detected in the mean line of the primary component at all observed phases. The longitudinal field, inferred from the LSD profiles, is variable, and preliminary modeling yields a probable period of about 8 d, changing sign during the rotational cycle. The variation appears to be sinusoidal, which would imply that the magnetic field likely has an important dipole component with a polar strength of at least 1.8~kG. The spectra show no clear profile variability or chemical peculiarity (particularly of He), suggesting that the photospheric properties of HD~200775 are not currently suitable for sustaining the abundance anomalies of Bp stars. 

\subsection{HD~190073}

HD~190073 is an A2 HAeBe star, showing a spectrum with a large number of emission lines. Like V380 Ori, many emission lines have superimposed narrow absorption cores, with a width similar to the photospheric lines with no emission. As reported by Catala et al. (in preparation), the emission components have a uniform width equal to 65~km/s. P Cyg profiles are exhibited by H$\alpha$ and other Balmer lines, with strong monthly variability. Acke \& Waelkens (2004) report no significant chemical peculiarity (Ca may be marginally enhanced). Pogodin et al. (2005) have reported complex structure in the Ca~{\sc ii}~H and K line profiles, and suggest that a magnetic field could account for the particular circumstellar structure of this star.

According to Catala et al. (in preparation), magnetic field has been measured from 8 ESPaDOnS spectra of this star, yielding LSD detections in each case. The Stokes $V$ signatures appear to be non-variable, and correspond to a constant longitudinal field of about 100~G. Like V380 Ori, HD 190073 probably has either a very long rotational period, is viewed nearly pole-on, or has magnetic and rotational axes that are nearly aligned.

\subsection{HD 139614, HD 144432 and HD 31648}

\begin{figure}
\centering
\includegraphics[width=6.0cm]{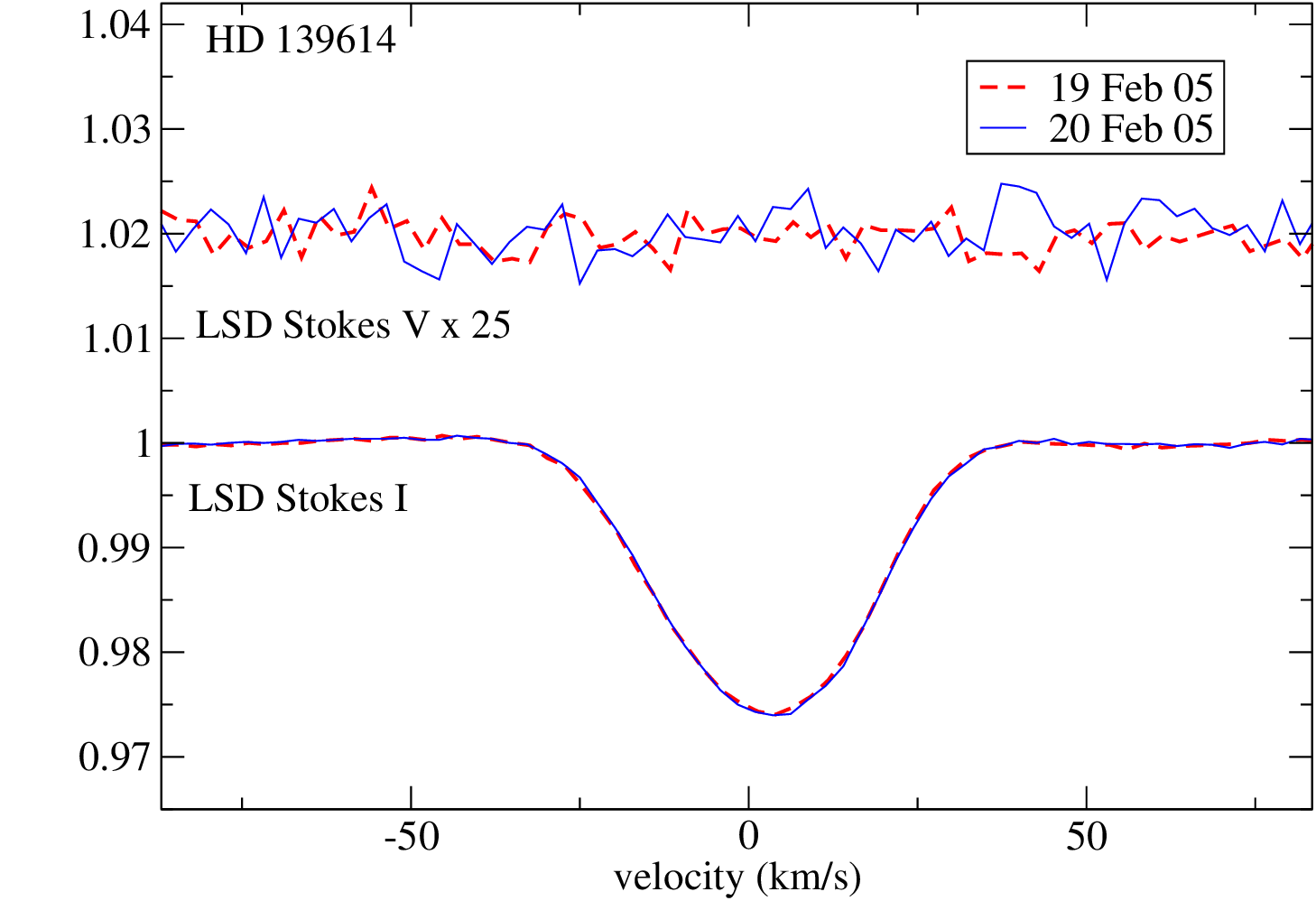}
\caption{Stokes $I$ and $V$ LSD profiles of HD 139614, derived from observations obtained on two separate nights. Note the absence of any Zeeman signatures associated with the absorption line. The inferred longitudinal field is weaker than about 25~G on both occasions. Adapted from Wade et al. (2005).}
\end{figure}

HD 139614, HD 144432 and HD 31648 are HAeBe stars in which weak magnetic field detections have been previously claimed by other authors. Each of these stars has been observed under good conditions with both FORS1 and with ESPaDOnS, and we have detected no magnetic field in any case. The example of HD 139614, observed twice with ESPaDOnS with no detection, is illustrated in Fig. 2.

\section{Conclusions}

Magnetic fields can be detected and measured in Herbig Ae/Be stars using methods of measurement currently available. From our initial survey results we have detected (and confirmed) 4 new magnetic PMS A and B stars. Unconfirmed indications of magnetic field are also obtained for two additional PMS stars. On the other hand, we are unable to confirm claims of magnetic field in 3 other Herbig Ae/Be stars.

Analysis of the magnetic field measurements leads us to conclude that the magnetic field characteristics of PMS stars of intermediate-mass are qualitatively and quantitatively similar to those of their PMS descendants: they demonstrate similar incidence (about 10\%), similar structure (primarily dipolar), and similar intensity (about 1~kG). Although chemical peculiarity cannot be confirmed in most of the magnetic Herbig Ae/Be stars, chemical spots are observed in the most evolved magnetic star of our sample. This strongly suggests that the magnetic Herbig Ae/Be stars and the magnetic Ap/Bp stars form an evolutionary sequence - that is, that the magnetic PMS stars that we have detected are the PMS progenitors of the Ap/Bp stars.


%


\end{document}